\documentclass[article,twocolumn,prd,superscriptaddress,preprintnumbers,nofootinbib]{revtex4-1}
\usepackage[utf8]{inputenc}
\usepackage{amsmath,amsthm,amssymb,mathtools,physics,cancel,mhchem,tikz}
\usepackage{graphicx,flexisym,float,mwe}
\usepackage{dcolumn}
\usepackage[hidelinks]{hyperref}
\hypersetup{colorlinks=true,
    linkcolor=blue,
    filecolor=blue,
    urlcolor=blue,
    citecolor=blue,
    }

\usepackage{caption}
\usepackage{subcaption}
\usepackage{adjustbox}
\usepackage{gensymb}
\usepackage{breqn}
\usepackage{braket}
\usepackage{enumitem}
\usepackage{gensymb}

\graphicspath{{./figures/}}

\begin{document}
\hfill \preprint{MI-HET-795}
\title{Probing the dark sector with nuclear transition photons}

\author{Bhaskar Dutta}
\email{dutta@physics.tamu.edu}

\author{Wei-Chih Huang}
\email{s104021230@tamu.edu}
\affiliation{Mitchell Institute for Fundamental Physics and Astronomy$,$ Department~ of ~Physics ~ and~ Astronomy$,$\\ Texas A$\&$M University$,$~College~ Station$,$ ~Texas ~77843$,$~ USA}

\author{Jayden L.~Newstead}
\email{jnewstead@unimelb.edu.au}
\affiliation{ARC Centre of Excellence for Dark Matter Particle Physics$,$ \\~School of Physics$,$~ The~ University~ of~ Melbourne$,$~ Victoria~ 3010$,$~ Australia}

\begin{abstract}
Here we present world-leading sensitivity to light ($< 170$ MeV) dark matter (DM) using beam-dump experiments. Dark sector particles produced during pion decay at accelerator beam-dumps can be detected via scattering in neutrino detectors. The decay of nuclei excited by the inelastic scattering of DM is an unexploited channel which has significantly better sensitivity than similar searches using the elastic scattering channel. We show that this channel is a powerful probe of DM by demonstrating sensitivity to the thermal relic abundance benchmark in a scalar DM dark-photon portal model. This is achieved through the use of existing data, obtained by the KARMEN experiment over two decades ago, which allow us to set world-leading constraints on this model over a wide mass range. With experimental improvements planned for the future, this technique will be able to probe the thermal relic benchmark for fermionic DM across a wide mass range.
\end{abstract}

\maketitle

Probes of dark-sector particles take many forms, including both direct and indirect DM searches. These searches were primarily targeted at constraining Weakly interacting DM candidates (WIMPs) \cite{Baudis:2013bba,
Vietze:2014vsa,Sahu:2020kwh,Klos:PRD2013,Arcadi:2019hrw}. However WIMP-like DM has not yet been detected  \cite{PhysRevLett.121.111302,PhysRevLett.123.251801,PhysRevD.103.063028} which has led to new DM models which expand the available solutions to the DM problem. Such models, built to circumvent past and present DM constraints, therefore require new methods for detection.

Light DM with a vector mediator, for example a dark photon, has been proposed in numerous studies as a viable DM candidate \cite{COHERENT:2021pvd, PhysRevD.84.075020, PhysRevLett.123.121801,PhysRevLett.113.171802,Dutta:2019nbn}. Previous searches have looked for the elastic scattering signature of DM in the detectors of pion decay-at-rest ($\pi$DAR) neutrino experiments, such as COHERENT \cite{coherent2018,Akimov:2021dab} at the Spallation Neutron Source (SNS) and Coherent CAPTAIN-Mills (CCM) at Los Alamos National Laboratory (LANL) \cite{CCM}. The relatively small size of the detectors and their higher background rates limit their sensitivity to DM. In this paper we perform a DM search via the inelastic channel, which has as a signal the narrow photon spectrum produced by the decay of excited nuclear states. Crucially, the larger energies deposited during inelastic scattering ensure that the sensitivity is not limited by detector thresholds, allowing the use of much larger higher-threshold detectors. Additionally, while the inelastic channel has a smaller cross section than the elastic channel, the narrow signal width permits a line search, significantly reducing the background. However, neutral current (NC) inelastic scattering of neutrinos remains an irreducible component of the background. This process was previously observed by KARMEN, a $\pi$DAR experiment at the ISIS neutron source~\cite{199815,KARMEN:1991vkr,Maschuw:1998qh}, which used the $^{12}$C($\nu$, $\nu'$)$^{12}$C$^*(1^+,1; 15.1$ MeV) reaction. This existing measurement is in agreement with the Standard Model prediction, therefore we can use it to place a constraint on DM causing additional carbon excitations. In the future this strategy can be used to search for DM in large beam-dump experiments such as the proposed PIP2-BD~\cite{Toups:2022yxs} with better sensitivities than the searches via elastic scattering.

Predicting the DM signal requires the same nuclear matrix elements used for neutrino NC inelastic reactions. Calculations of neutrino nucleus scattering cross sections have a long history~\cite{DONNELLY19748} (see \cite{EJIRI20191} for a review). While some reactions, like the carbon one measured by KARMEN, are well studied, most have not been. Predicting the total observable signal remains computationally challenging since it includes many final states. Inclusive methods have been applied to argon~\cite{VanDessel:2019atx}, however they do not provide information about the relative populations of the final states, making it difficult to predict the signal spectrum. Recent calculations~\cite{inelastic2022} and experimental data~\cite{Tornow:2022kmo} confirm that Gamow-Teller (GT) transitions dominate the cross section in this regime. Considering only GT transitions simplifies the calculation of the DM and neutrino spectra, allowing us to perform this search for DM through the inelastic channel with world-leading sensitivity to light DM.

{\bf{\emph{Beam-dump experiments }-}} High-energy proton beams impinging on dense targets (e.g. mercury or tungsten) produce large numbers of pions. The $\pi^-$ are captured by nuclei before they decay and the $\pi^-$ are stopped by the target. The stopped pions efficiently decay to muons, producing a well understood spectrum of neutrinos. These neutrinos have been used to study their low-energy interactions with nuclei. For example, coherent elastic neutrino-nucleus scattering (CE$\nu$NS) and NC inelastic scattering were first observed in this way by the COHERENT~\cite{Akimov:2017ade} and KARMEN~\cite{KARMEN:1991vkr} experiments, respectively. The proton collisions could also produce a large flux of DM, primarily through the decay of pions, which could then scatter in the neutrino detectors. The DM flux is relativistic, overcoming the kinematic suppression which severely limits the sensitivity of DM direct detection experiments to sub-GeV DM.

\begin{table*}[th]
    \centering
    \caption{Specifications of the experiments and detectors ($^\dagger$ indicates proposed)}
    \begin{tabular}{c|c c c| l c c c c c}
    \hline \hline
    Experiment & $E_{\rm beam}$ & POT & Target & Detector: & & & & \\
     & [GeV] & [yr$^{-1}$] & & material & mass & distance & angle & runtime & $E_r^{\rm th}$\\
   \hline
   KARMEN~\cite{Reichenbacher:2005nc} & 0.8 & $ 1.16 \times 10^{22} $ & Ta & CH$_2$ & 56~t & 17.7~m & 100$^\circ$ & 4 years & 10 MeV\\ 
   COHERENT$^\dagger$ & {1}  & {$6.0\times 10^{23}$} & {Hg} & NaI[Tl] & 3.5~t & 22~m & 120$^\circ$ & 3 year & $\sim$few~keV\\
   \cite{Akimov:2017ade, COHERENT:2019kwz, coherent2018} & & & & & & & & & \\
   CCM$^\dagger$~\cite{CCM, CCM:2021leg} & 0.8 & $7.5\times 10^{21}$ & W & Ar & 7~t & 20~m & 90$^\circ$ & 3 years & 25~keV \\
   PIP2-BD$^\dagger$~\cite{Toups:2022yxs} & 2 & $9.9\times 10^{22}$ & C & Ar & 100~t & 15~m & N/A & 5 years & 20~keV \\
    \hline \hline
    \end{tabular}
    \label{tab:expspec}
\end{table*}

The KArlsruhe Rutherford Medium Energy Neutrino (KARMEN) experiment was located at the ISIS neutron source, which has an 800~MeV proton beam, pulsed at 50 Hz, directed into a tantalum beam stop. We make use of the $^{12}$C($\nu_\mu$, $\nu_\mu'$)$^{12}$C$^*$ reaction analysis which included 4.65$\times 10^{22}$ protons-on-target (POT)~\cite{199815}. The detector was a liquid scintillator calorimeter with total mass 56 tonne ($\sim 10^{30}$ $^{12}$C nuclei). In this analysis 86$\pm 15$ $^{12}$C $(\nu_\mu,\nu_\mu')^{12}\text{C}^*$ events were observed.

COHERENT, based at the SNS, uses a 1 GeV proton beam (width 0.6 $\mu s$), pulsed at 60 Hz, which impinges on a mercury target at a rate of $8.8 \times 10^{15}$ POT/s. The COHERENT program runs (or have plans for) six detectors with different nuclear targets in the so-called ``Neutrino Alley", of which we are interested in NaI.

The current NaI[Th] detector has mass of 185 kg, a threshold of roughly 900 keV, is located 22 m away from the target. It was recently decommissioned to be replaced by NaIvETe, which has a mass of 3.5 tonne~\cite{2022NaIvETe}. The threshold is expected to be a few keV$_{\mathrm{ee}}$. The background is approximately flat and O(100) in the total exposure.

CCM at LANL make use of a 0.8 GeV proton beam ($0.29 \mu s$ wide, 20 Hz frequency) impinging on a tungsten target, which gives $5.6\times10^{14}$ POT/s. Currently they are operating a liquid argon (LAr) detector with a 7 tonne fiducial volume, located 20 m away from the target with a 25 keV threshold.

PIP2-BD at Fermilab~\cite{Toups:2022yxs} will use a 2 GeV proton beam ($2 \mu s$ wide), 120 Hz, impinging a light target such as carbon. A 100 tonne LAr detector with 20 keV threshold will be located at different distances (15 m or 30 m) from the target and at different angles.

The ratio of $\pi^-$ production to POT is 0.0457 for COHERENT, 0.0259 for CCM, and 0.233 for PIP2-BD, while the ratio for $\pi^0$ production is 0.1048 for COHERENT, 0.0633 for CCM~\cite{Dutta:2020vop}, 0.322 for PIP2-BD, and 0.0448 for KARMEN~\cite{Plischke:2001ii}. In table~\ref{tab:expspec} the key specifications of these experiments are summarized. For our inelastic search channels, we will utilize the sensitivity of these experiments to $\gtrsim$ MeV energy $\gamma$-rays.

\begin{figure}[t]
    \centering
    \includegraphics[width=\columnwidth]{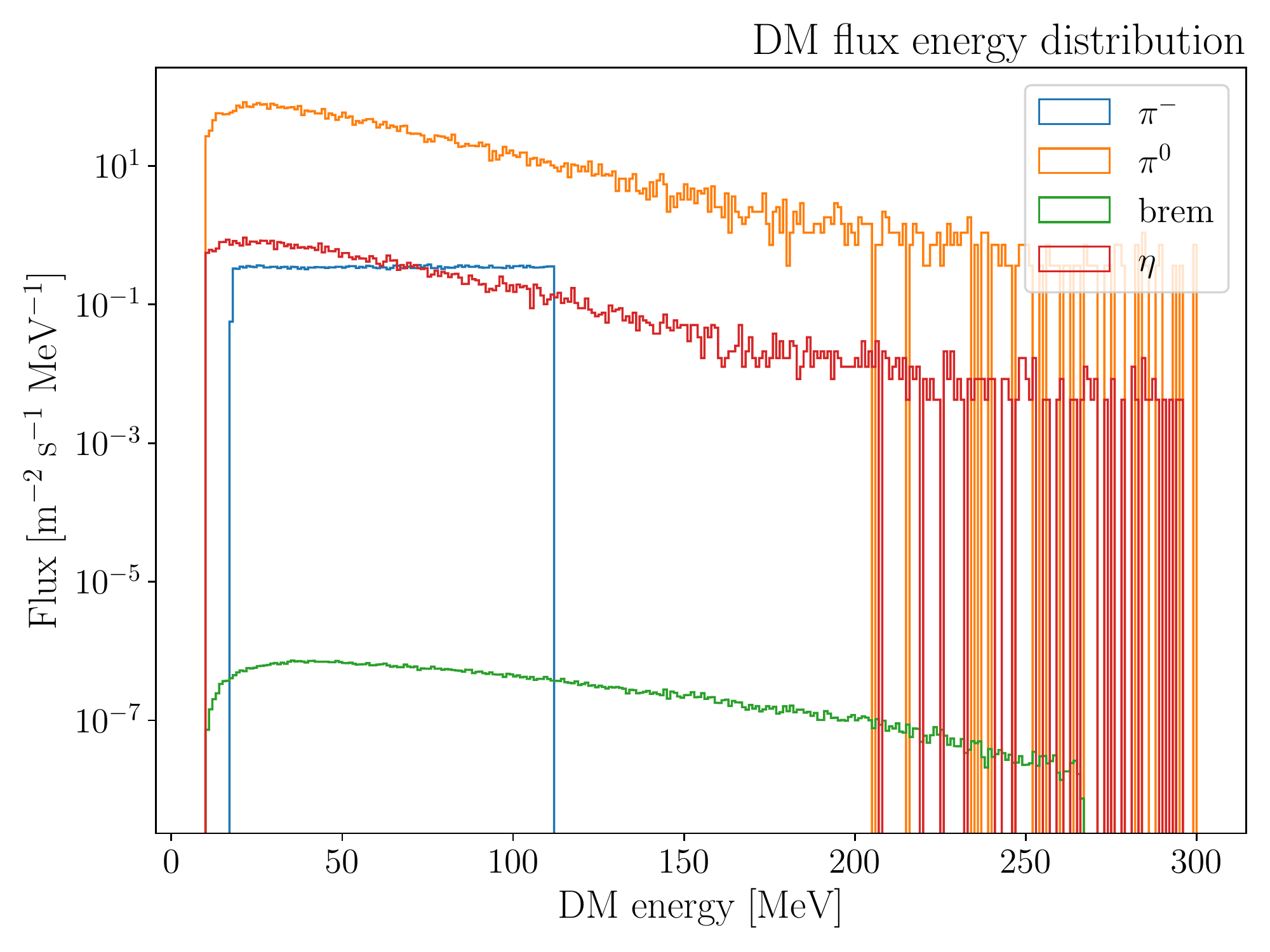}
        \caption{The contributions to the DM flux at CCM LAr detector, where we have simulated with $m_{A'}=3m_\chi=30$ MeV and $\epsilon=10^{-4}$.}
    \label{fig:dmflux}
\end{figure}

{\bf{\emph{Light DM scattering}-}} A theoretically appealing thermal DM model can be realized via a minimal extension to the Standard Model (SM) where a light DM particle is coupled to quarks via a dark photon ($A'$). A small coupling to quarks is achieved through the $A'$ kinetically mixing with the SM photon~\cite{Pospelov:2007mp,Chun:2010ve,Batell:2014yra}. The interaction Lagrangian for fermionic, $\chi$, and scalar, $\phi$, DM coupled to the SM via the dark photon is expressed as
\begin{eqnarray}
    \mathcal{L}_f &\supset& g_D A'_\mu \bar \chi \gamma^\mu \chi + e \epsilon Q_q A'_\mu \bar q \gamma^\mu q \\
    \mathcal{L}_s &\supset& \lvert D_\mu \phi \rvert^2 + e \epsilon Q_q A'_\mu \bar q \gamma^\mu q
    \nonumber
    \label{eq:lag}
\end{eqnarray}
where $g_D$ is the dark coupling constant, $\epsilon$ is the mixing parameter, $Q_q$ is quark's electric charge. The dark photon is produced in any process with SM photon production. For example, they can be produced through pion capture, pion decay and photons emerging from cascades (bremsstrahlung):
\begin{eqnarray}
    \pi^- + p &\rightarrow& n + A' \\
    \pi^0 &\rightarrow& \gamma + A' \nonumber \\
    \eta^0 &\rightarrow& \gamma + A' \nonumber \\
    e^{\pm*} &\rightarrow& e^\pm + A' \nonumber
\end{eqnarray}
In a light DM scenario where $m_\chi < m_{A'}$, the dark photons then decay to DM: $A' \rightarrow \chi\bar{\chi}$. We assume $A'$ decays in flight to a pair of DM particles immediately after it is produced ($\lesssim 10^{-10}$ ns). Previously, GEANT4 (\cite{GEANT4}) has been employed to simulate the DM spectra from mesons and bremsstrahlung~\cite{Dutta:2020vop}. Fig.~\ref{fig:dmflux} shows a sample simulated DM flux at CCM's LAr detector, where $m_{A'}=3m_\chi=30$ MeV and $\epsilon=10^{-4}$. Due to its higher production ratio at $\sim$ GeV energies, $\pi^0$ decay dominates all other production channels.  Similar fluxes can be obtained at COHERENT, KARMEN and PIP2-BD.

The produced DM then propagates to the detectors where it may scatter from the detector nuclei, producing nuclear recoils and excitations. At low momentum transfer the inelastic scattering cross section is dominated by GT transitions (described by the operator $\frac{1}{2}\hat{\sigma_i} \hat{\tau_0}$) \cite{inelastic2022}. Thus to a good approximation the inelastic cross section to a given final state $J_f$ is:
\begin{align}
\label{eq:DMcs}
    \frac{d\sigma^{DM}_{inel}}{d\cos\theta} &= \frac{2e^2\epsilon^2 g_D^2 {E'}_\chi {p'}_\chi}{(2m_N E_r + m_{A'}^2 -\Delta E^2)^2} \frac{1}{2\pi} \frac{4\pi}{2J+1} \\
    &\times \sum\limits_{s_i,s_f} \vec{l} \cdot \vec{l}^* \frac{g_A^2}{12\pi}
    |\langle J_f|| \sum_{i=1}^A \frac{1}{2}\hat{\sigma_i} \hat{\tau_0}|| J_i\rangle|^2 \nonumber
\end{align}
where $\Delta E$, $m_N$, and $J$ are the excitation energy, nuclear mass and spin, respectively, and the axial coupling constant is $g_A=1.27$~\cite{PDG2000}. 
The DM currents, $\vec{l}$, depend on the DM spin under consideration. Here we treat both fermionic and scalar DM. After spin sums the current term is given by:
\begin{eqnarray}
    \hspace*{-4mm}\sum\limits_{s_i,s_f} \left(\vec{l} \cdot \vec{l}^*\right)_f &=& 3- {1\over 4E_\chi {E'}_\chi} \left[ 2 \left(p_\chi^2 + {p'}_\chi^2-2m_NE_r \right) + 3m_\chi^2 \right] \nonumber \\
     \hspace*{-4mm}\sum\limits_{s_i,s_f} \left(\vec{l} \cdot \vec{l}^*\right)_s &=& {1\over 2E_\phi {E'}_\phi} \left(p_\phi^2 + {p'}_\phi^2-2m_NE_r \right)
    \nonumber
\end{eqnarray}
This induces a factor of $\sim2$ difference in cross section between the fermionic and scalar DM, i.e. $\sigma^{DM}_f \sim 2 \sigma^{DM}_s$. 

Due to the effect of coherency, CE$\nu$NS cross sections are much larger. However the only observable signal is the nuclear recoil which can be challenging to detect at keV energies and are subject to large backgrounds. Relativistic light DM and neutrinos with $E=10-100$~MeV can excite nuclear states up to 15-30 MeV, depending on the target~\cite{inelastic2022}. Some of these states will have enough energy to decay via particle emission. For argon, the neutron emission threshold of $\sim 10$ MeV~\cite{SUTTON1983415} and angular-momentum barrier should result in a photon emission branching ratio close to 1 for all decays from $J=1^+$ states below $\sim11$ MeV. For sodium the proton emission threshold is around 8.8 MeV and so the high end of the region of interest may be affected. For iodine the threshold is also around 9 MeV but since the GT strength is negligible, it will not affect our results.

{\bf{\emph{GT transition strengths}-}}
To calculate the DM-nucleus cross section in Eq.~(\ref{eq:DMcs}) we must evaluate the strengths of the relevant GT transition. For $^{12}$C we make use of existing theory predictions which are in good agreement with data~\cite{199815}. We use the mean theory prediction for the flux-averaged cross section of the $^{12}\text{C}(\nu,\nu)^{12}\text{C}^*$ reaction to compute the GT strength (giving $\text{B(GT)}=0.255\pm0.021$) and use this in subsequent calculations.

\begin{figure}[tb]
    \centering
    \includegraphics[width=\columnwidth]{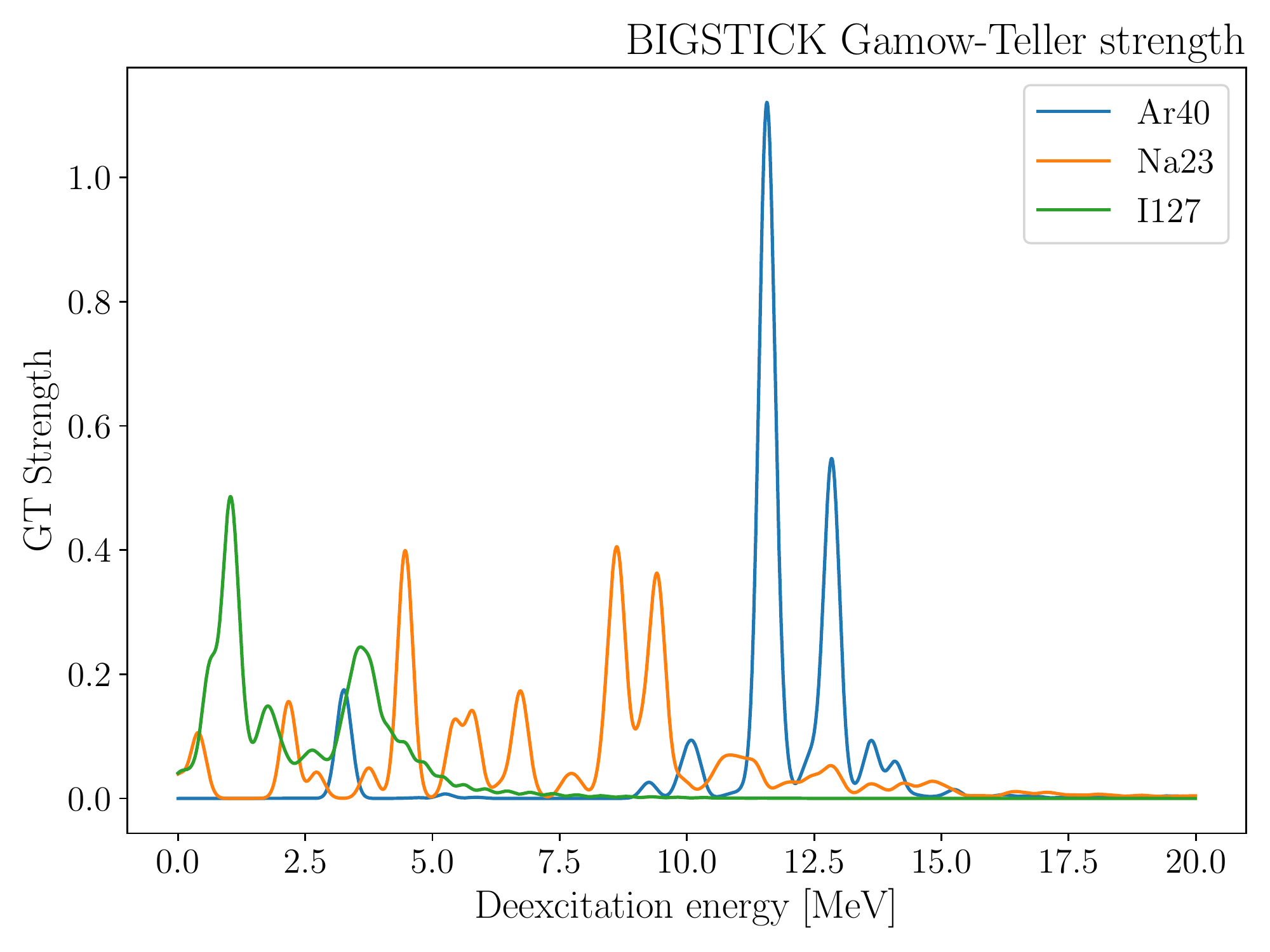}\\
    \caption{The GT strength convolved with a 150~keV width Gaussian for $^{40}$Ar, $^{23}$Na, and $^{127}$I.}
    \label{fig:strengths}
\end{figure}

For the remaining nuclear targets, $^{23}$Na, $^{40}$Ar, and $^{127}$I, we use the nuclear shell model code BIGSTICK to compute the strengths~\cite{Johnson:2018hrx,Johnson:2013bna}. The $^{23}$Na calculation is relatively simple because of its small number of valence nucleons and the small model space. There are 3 valence protons and 4 valence neutrons in {\it sd} orbits ($0d_{5/2}$, $1s_{1/2}$, $0d_{3/2}$). We use the {\it USDB} interaction for $^{23}$Na~\cite{USD,USDB}. The calculation for $^{40}$Ar is more challenging as the protons and neutrons are in different model spaces. The valence protons are in {\it sd} orbits ($0d_{5/2}$, $1s_{1/2}$, $0d_{3/2}$), while the valence neutrons are in {\it pf} orbits ($0f_{7/2}$, $1p_{3/2}$, $0f_{5/2}$, $1p_{1/2}$). We truncate {\it sdpf} space to reduce the computational workload, restricting the maximum number of protons excited to 4. Neutrons are constrained to the {\it pf} orbits. We use the {\it SDPF-NR} interaction \cite{Prados:PRC2007,Nowacki:PRC2009,Nummela:PRC2001} for $^{40}$Ar. For $^{127}$I, the model space is $0g_{7/2}$, $1d_{5/2}$, $0h_{11/2}$, $1d_{3/2}$ and $2s_{1/2}$ and we adopt the {\it jj55pna} interaction \cite{Brown:PRC2001}. 


Fig.~\ref{fig:strengths} shows the computed strength functions for the GT operator for our 3 nuclear targets, convolved with a 150 keV width Gaussian (giving energy resolution $\sim$15\%). Following \cite{Tornow:2022kmo} we scale the total GT strengths for argon in the 4-11 MeV range to match the experimental data of B(M1) = 0.651 $\mu_N^2$. While we don't perform the full multipole calculation, the resulting cross section is well approximated by the GT transition for low-energy, where our flux is concentrated~\cite{inelastic2022,Tornow:2022kmo}. The same procedure could be carried out for sodium and iodine if the data were available, though it is unlikely to make a large difference for this analysis. This is because prior comparisons with data for the charged-current reactions show agreement for sodium using the USDB model~\cite{PhysRevC.97.024310} and iodine does not contribute much to the total GT strength in the region of interest.

{\bf{\emph{Sensitivity}-}} Inelastic DM-nucleus scattering produces a small nuclear recoil and a subsequent cascade of deexcitation $\gamma$-rays. The cascade energy $\sim$MeV will dwarf the nuclear recoil energy $\sim$keV and therefore we ignore the contribution of the latter. Since the half-life of the decay cascade is extremely short (picosecond or even femtosecond level), we will treat the deexcitation process as a single energy deposition completely contained within the detector. This is a reasonable approximation for large argon detectors and KARMEN, but is less applicable to smaller detectors. A detailed analysis including detector geometry could account for partial detection of decay products but is beyond the scope of the present work.

The expected number of events can be computed from
\begin{equation}
    N = \frac{{\mathrm{exposure}}}{m_d} \times \int \sigma(E_\chi) \nonumber \frac{d\Phi}{dE_\chi} dE_\chi
\end{equation}
where the exposure $=\text{running time} \times \text{detector mass}$, $m_d$ is the mass of a single molecule of the detector material, and $\frac{d\Phi}{dE_\chi}$ is the DM energy flux. We assume that the detectors have 100\% detection efficiency (except for KARMEN which had a 20\% efficiency) and that all energy depositions are above threshold.

\begin{figure}[tbh]
    \centering
    \includegraphics[width=\columnwidth]{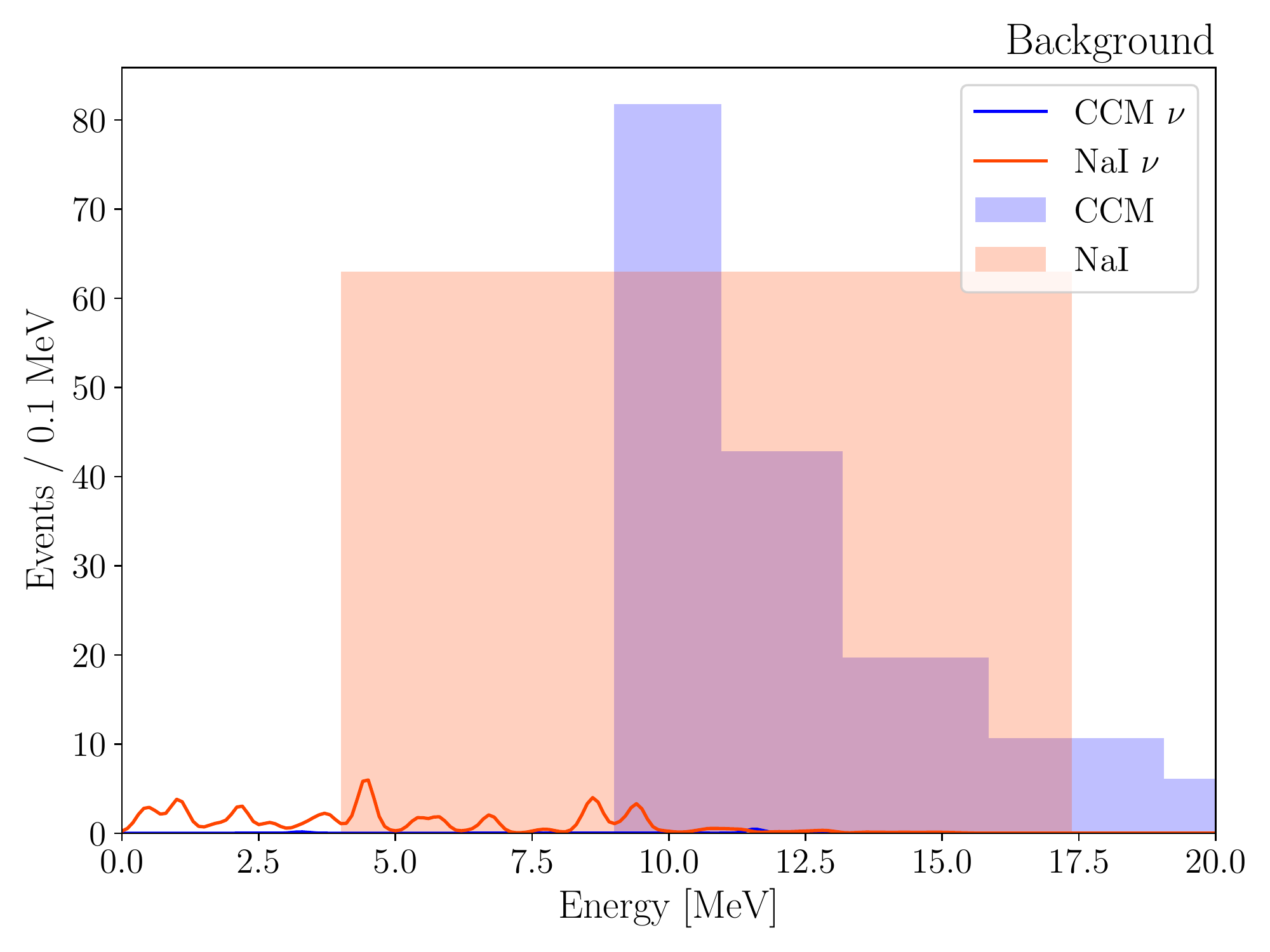}\\
    \caption{Background spectrum expected for the full exposures of the COHERENT NaI (3.5t) and CCM LAr detectors (solid) and the inelastic neutrino scattering contributions (curves). The timing cut and strength scaling are applied.}
    \label{fig:bkg}
\end{figure}

Backgrounds for the full exposures of COHERENT NaI (3.5t) and CCM LAr are expected to be $O(100)$ events in the region-of-interest $E= 1-100$ MeV. For CCM we take the background distribution from the engineering run in  and assume the science run will achieve the expected improvement of 100 times lower background rate~\cite{CCM:2021lhc}. For COHERENT NaI there is no available data and so we assume a flat background with $\sim O(100)$ events across the region-of-interest. For KARMEN we take the background-subtracted $^{12}$C($\nu_\mu$, $\nu_\mu'$)$^{12}$C$^*$ events ($86\pm15$) as background to a DM signal (since any DM events would have mimicked these). To show the most optimistic projection, for PIP2-BD the background is assumed to due to neutrinos only.

Fig.~\ref{fig:bkg} shows the assumed background spectra for the two detectors. Lower energy cuts of $4$ MeV and $9$ MeV are applied to the NaI and LAr detectors respectively since we are using the lines located above 4 MeV for NaI and 9 MeV for LAr. The detector background decrease for higher energy lines. Beyond these detector backgrounds, we also include deexcitation photons produced from inelastic $\nu$-nucleus scattering. The sensitivity of this line-based search can be improved by using a single line and a detector with high energy-resolution. 

The inelastic $\nu_e$ and $\bar{\nu}_\mu$ background can be reduced by applying a prompt window timing cut, here taken to be within 150 ns of the beam arrival. This was demonstrated in KARMEN~\cite{199815} and CCM~\cite{CCM:2021lhc}, and we assume this cut applies to COHERENT as well. With such a cut, the $\nu$ background will only be due to prompt $\nu_\mu$ from pion decay giving rise to neutral current events. 
The DM produced from pion and eta decay propagates relativistically to the detector and is unaffected by the cut. To compute the inelastic $\nu$-nucleus cross sections and background rates we make use of the results from~\cite{inelastic2022}.

\begin{figure}[tbh]
    \centering
    \includegraphics[width=\columnwidth]{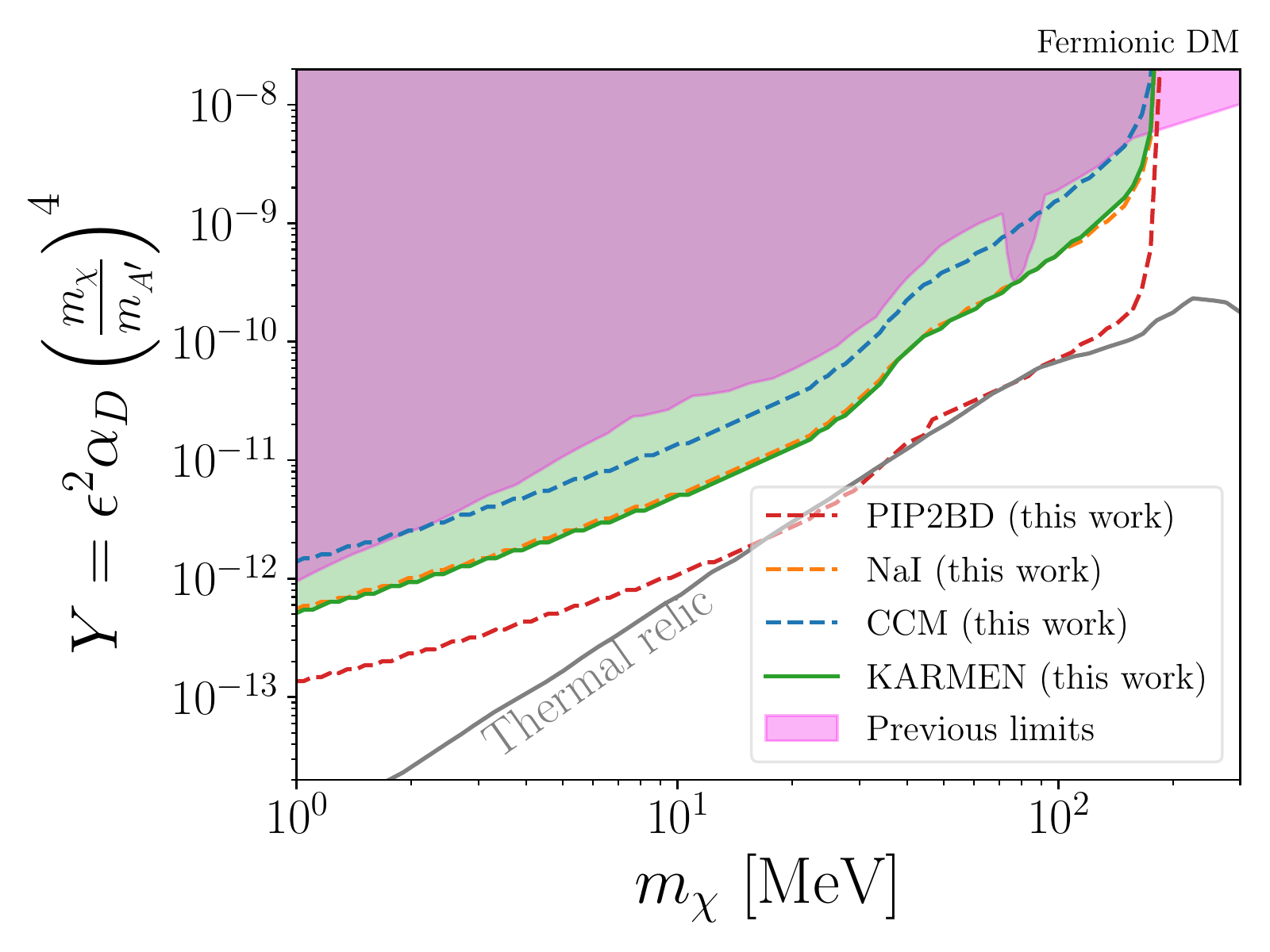}\\
    \includegraphics[width=\columnwidth]{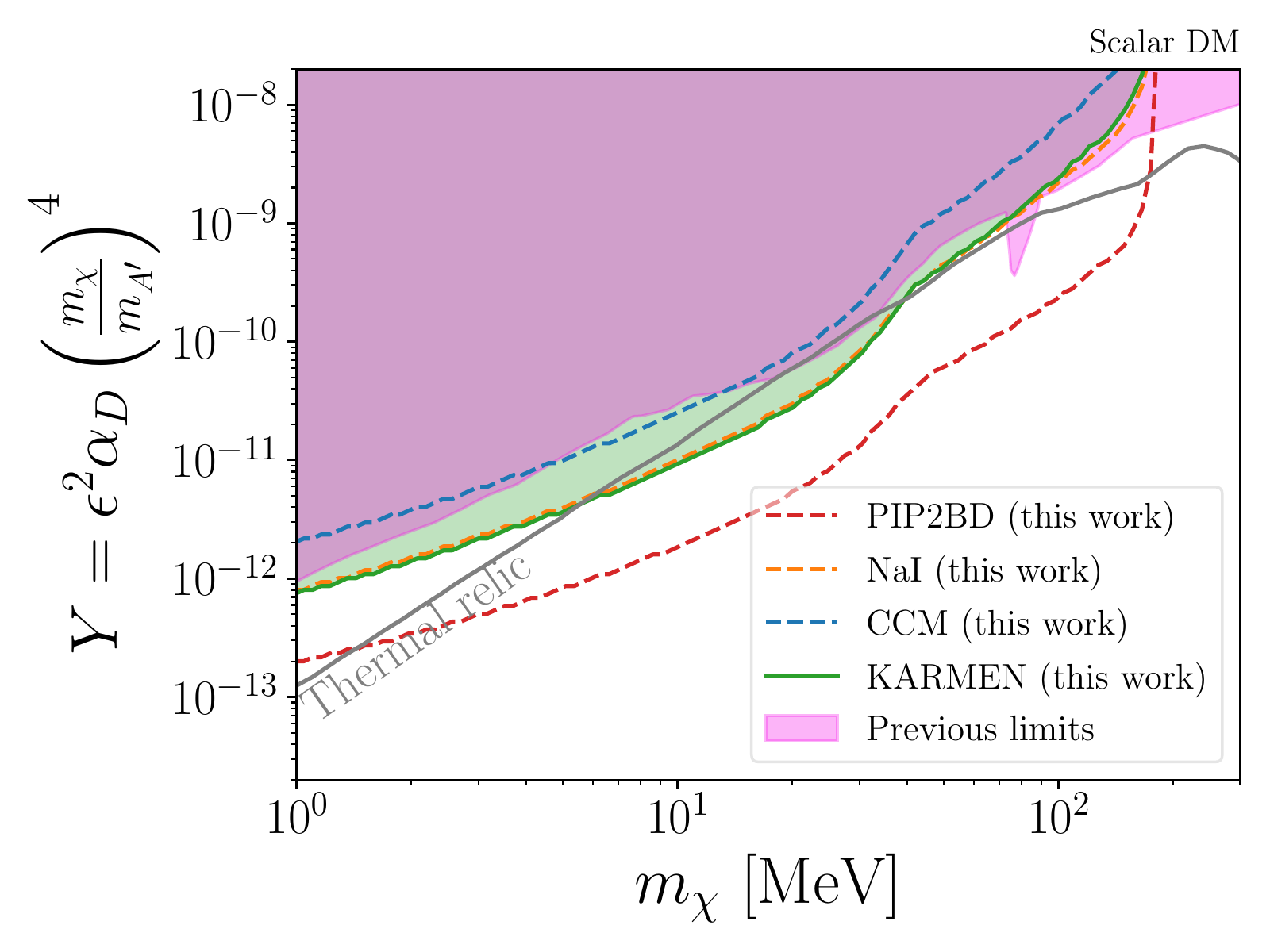}\\
    \caption{The 90\% CL exclusion bounds on fermionic (top) and scalar (bottom) DM for various experiments. The previous bounds are shaded in magenta, while the green shaded regions denote the new bounds derived in this work. Dashed curves denote projections for future experiments (this work).}
    \label{fig:sen}
\end{figure}

We investigate the DM parameter space fixing the mass ratio $m_{A'}/m_\chi = 3$ and $g_D=\sqrt{2\pi}$. The derived exclusion bounds for KARMEN and projected sensitivity of the future COHERENT, CCM and PIP2-BD experiments are shown in Fig.~\ref{fig:sen} where we have used a $\chi^2$ test at 90\%CL. The magenta shaded region shows the existing limits from various elastic scattering searches~\cite{COHERENT:2021pvd,LSND:2001PRD,CCM:2021leg,MiniBoone:2018PRD,NA64:Gninenko_2021,E137:1988PRD,PhysRevD.104.L091701}, the green shaded region shows the new limit based on our inelastic nuclear scattering calculation using the existing KARMEN data and the dashed curves show the projected sensitivity that can be obtained with inelastic scattering searches. The gray solid curve shows the combination of parameter values which achieve the correct thermal relic abundance of DM.

Our result provides the strongest current bound on this DM model across a wide range of DM masses, $m_\chi\sim 1$-$100$ MeV. For the case of scalar DM this result probes the thermal relic benchmark over a wide range of masses, $m_\chi\sim 5$-$40$ MeV. Future experiments greatly extend this range and even provide sensitivity to the themal relic benchmark for the fermionic DM scenario, which has thus far been out of reach.

The lines have a kink at around $m_\chi=45$ MeV due to the dominant source of DM production ($\pi^0$ decay) becoming closed when $m_{A'} = 3m_\chi = 135 \text{MeV} \approx m_{\pi^0}$. The $\eta^0$ decay flux then becomes the dominant source of DM for $m_\chi \ge 45$ MeV. This affects both elastic and inelastic channels and so the inelastic channel remains the most sensitive across the whole mass range.

The sensitivity of the searches using the elastic channels flattens for $m_\chi \leq 30$ MeV due to detector thresholds. The inelastic channel reach, however, continues to become stronger as the mass of DM decreases since the deexcitation lines are in the MeV region. The sensitivity shown for a 100 tonne LAr detector can be improved further with increased POT, which is the plan for PIP2-BD at Fermilab~\cite{Toups:2022yxs}.

{\bf{\emph{Conclusions}-}} We have performed the first investigation of inelastic nucleus scattering as a probe of dark sector physics. The NC inelastic channel has lower experimental background and much higher energy compared to the elastic channel. These characteristics make the inelastic channel the most sensitive probe of the light DM models considered here using existing and ongoing experiments. In this initial investigation, we made use of GT transitions in carbon, argon, sodium, and iodine nuclei relevant to past, present and future $\pi$DAR experiments: KARMEN, COHERENT, CCM and PIP2-BD. Using the inelastic channel, we find that the KARMEN experiment has world-leading sensitivity to DM produced from the decay of dark photons, reaching the thermal relic benchmark for a wide range of masses.

\begin{acknowledgments}
We thank Calvin W.~Johnson for detailed discussions and help with BIGSTICK, which made this work possible. We also thank Sean Finch, Gordan Krnjaic, Louis Strigari, Vishvas Pandey, Dan Pershy, Surjeet Rajendran,  Matthew Toups, Kate Scholberg, Richard Van de Water, for valuable discussions and comments. The work of BD and WH are supported in part by the DOE Grant No. DE-SC0010813.  JLN is supported by the Australian Research Council through the ARC Centre of Excellence for Dark Matter Particle Physics, CE200100008.
\end{acknowledgments}

\bibliographystyle{bibi}
\bibliography{refs}

\end{document}